\documentclass[10pt,letterpaper]{article}
\usepackage[top=0.85in,left=2.75in,footskip=0.75in]{geometry}

\usepackage{amsmath,amssymb}

\usepackage{changepage}

\usepackage[utf8x]{inputenc}

\usepackage{textcomp,marvosym}

\usepackage{cite}

\usepackage{nameref,hyperref}

\usepackage[right]{lineno}

\usepackage{microtype}
\DisableLigatures[f]{encoding = *, family = * }

\usepackage[table]{xcolor}

\usepackage{array}

\newcolumntype{+}{!{\vrule width 2pt}}

\newlength\savedwidth

\raggedright
\usepackage[aboveskip=1pt,labelfont=bf,labelsep=period,justification=raggedright,singlelinecheck=off]{caption}

\makeatletter
\renewcommand{\@biblabel}[1]{\quad#1.}
\makeatother

\usepackage{lastpage,fancyhdr,graphicx}
\usepackage{epstopdf}
\fancyheadoffset[L]{2.25in}
\fancyfootoffset[L]{2.25in}
\begin{document}
\vspace*{0.2in}

\begin{flushleft}
{\Large
\textbf\newline{Efficacy versus abundancy: comparing vaccination schemes} 
}
\newline
\\
Omar El Deeb* \textsuperscript{1,2 \Yinyang} and 
Maya Jalloul \textsuperscript{3 \Yinyang}

\bigskip
\textbf{1} Department of Natural Sciences, Lebanese American University, Beirut, Lebanon
\\
\textbf{2} Faculty of Technology, Lebanese University, Aabey, Lebanon
\\
\textbf{3} Department of Economics, Lebanese American University, Beirut, Lebanon
\\
\bigskip

%
%
\Yinyang These authors contributed equally to this work.

* omar.eldeeb@lau.edu.lb

\end{flushleft}
\section*{Abstract}

\begin{abstract}
We introduce a novel compartmental model accounting for the effects of vaccine efficacy, deployment rates and timing of initiation of deployment.
We simulate different scenarios and initial conditions, and we find that higher abundancy and rate of deployment of low efficacy vaccines lowers the cumulative number of deaths in comparison to slower deployment of high efficacy vaccines.
 We also forecast that, at the same daily deployment rate, the earlier introduction of vaccination schemes with lower efficacy would also lower the number of deaths with respect to a delayed introduction of high efficacy vaccines, which can however, still achieve lower numbers of infections and better herd immunity.

\end{abstract}
keywords: Compartmental model; Vaccine; COVID-19; Epidemiology.

\section{Introduction}

Humanity has been struggling with viral pandemics and infectious diseases
that caused great catastrophes throughout the entire history of mankind.
Since the Athenian, Antonine and Justinian Plagues to the Black Death,
Spanish Flu, Cholera, and Smallpox up until the HIV pandemic, SARS,
Swine flu, Ebola outbreak and, most recently, the Coronavirus infection,
pandemics have been a major source of disease, death, economic crises
and political turmoils \cite{History1,History2,History3}. Nevertheless,
those outbreaks also lead to major discoveries and advancements in
sciences and public health, especially in medicine, pharmaceuticals,
vaccines and development of public policies \cite{HistMed,HistVac1,HistVac2}.

The Coronavirus disease (COVID-19) associated to the severe acute
respiratory syndrome coronavirus 2 (SARS-CoV-2) continues as a main
cause of hospitalization and death and as a main public health risk
since the first case was registered in Wuhan, China in December 2019
\cite{First}. It was declared as a global pandemic on March 11, 2020
\cite{pandemic} after spreading from China into several Asian and
European countries and consequently into tens of countries across
the world. By December 22, 2021, the world has suffered about 278 million
registered infections and 5.4 million deaths associated to COVID-19
in 224 different countries, territories and entities \cite{Worldometer}.
Health systems were put under enormous pressure, and the consequent
mitigation measures and associated closures and lockdowns taken across
the world for lengthy periods of time contributed to a global economic
slowdown and recession in several countries \cite{Econ1,Econ2}, in
addition to a heavy negative impact on education, employment, tourism
and social activities \cite{Effects1,Effects2,Effects3} exacerbating
existing political discontent and stirring political unrest \cite{Political}.

Similarly to the other respiratory pathogens, airborne transmission
of SARS-CoV-2 occurs by inhaling droplets loaded with the virus emitted
by infectious people. Infection can also occur through mucous membranes
like the eyes, nose or mouth \cite{Transmission}.

Many variants of the virus emerged with an increased risk to global
public health. The World Health Organization (WHO) characterized variants
of interest (VOI) and variants of concern (VOC) to track and monitor
\cite{Variants}. The main variants of concern have stronger capabilities
of transmissibility and health risks, and they are labeled by the WHO
as the Alpha, Beta, Gamma, Delta and Omicron variants which were first documented
in the United Kingdom, South Africa, Brazil and India, respectively \cite{Delta}.
The recent emergence of variants creates a major cause
of concern since they can lead to an epidemic rebound especially with
the possibility of vaccine resisting and the emergence of deadlier
and/or more transmissible future variants. Increased viral transmission
means higher probability of emergence of variants as well \cite{Variants2}.

Several types of vaccines were developed and approved by global (WHO)
or by national health agencies, using different techniques like the
viral vector vaccines (Oxford/AstraZeneca, Sputnik V/Gamaleya, Janssen),
genetic vaccines (Pfizer/BioNTech, Moderna), inactivated vaccines
(Sinovac, Sinopharm, Bharat) and protein vaccines (Novavax, Sanofi)
\cite{Vaccines}. By December 22, 2021, a total of about 8.85 billion
doses of those vaccines were administered with main concentration
in China, USA, India and Europe \cite{DataVac}. Other vaccines, inhaled
aerosol dry vaccines and antiviral drugs are under extensive research
and development \cite{OtherMed1,OtherMed2,OtherMed3}. Antibody therapies
are under development as well and they offer an effective treatment
for the most seriously ill patients, but they remain expensive and
in short supply \cite{Antibody}.

A global roll-out of vaccines is needed to guarantee a swift elimination
of the infection, and for this end vaccines need to be available,
affordable, and accessible at a global scale. In high-income countries,
large stocks were produced and purchased, combined with active logistic
and public health resources, but in low-income countries, vaccination
is still slow mainly due to insufficiency of vaccines. The WHO has
called for more equity and stronger support for its COVAX initiative
in order to supply wider vaccine access for poor countries, but statistics
show that the gap remains deep as many rich countries have already
bought multiple times of doses needed per person \cite{Distribution}.
A similar gap also arises on the national level as health illiteracy,
religious beliefs and sometimes political partisanship are slowing
down vaccinations and creating irregular immunity rates among different
regions or groups \cite{Asymmetric}. As such, the global efforts
to stop the spread remains far from successful, and the priority should
be focused to optimize the use, deployment and timing of available
vaccines, whether those recognized by the WHO or by national health
agencies across the globe in order to prevent further deaths as well
as emergence of new variants.

The spread patterns of infectious diseases have been and are being
actively studied and simulated using several mathematical and computational
models in the interdisciplinary fields of health economics, physical biology and medical physics, applied mathematics and epidemiology .
A wide variety of techniques were employed ranging from compartmental
models, Agent-Based models (ABM), spatio-temporal analysis to data-driven
analysis and artificial intelligence \cite{Modelling1,Modelling2}.
The first compartmental model was the famous Kermack--McKendrick
Susceptible-Infectious-Removed (SIR) model which divides the total
population under study into three compartments where agents would
move from one compartment to another upon infection or removal through
recovery or death \cite{Kermack}. Numerous variations of this model
were introduced to account for characteristics and dynamics of different
diseases as well as in simulating similar interactions and spreads
in the social and behavioral sciences \cite{Comp1,Comp2,Comp3}. The
research on COVID-19 spread extensively employed SIR models that were
improved to account for other effects like exposure, travel, quarantine,
vaccination, public measures and many other alterations to best describe
the underlying features of the contagion \cite{sir1,sir2,sir3,sir4,sir5},
as well as general and country specific ABMs, spatio-temporal and
data driven studies \cite{ABM1,ABM2,ST1,ST2,ST3, scr1, scr2, RSR, PL1}.

However, with the mass introduction of various globally and nationally
recognized vaccines with varying efficacies, and with the varying
degrees of availability and logistic capabilities across the globe,
an important question arises for decision makers: which vaccine to
choose to deploy in a certain country from the available options in
that country, given their efficacies, their availability timeline
and expected deployment rate of each. This is a major concern for
public health officials aiming at minimizing the number of deaths
and/or the number of active infections, that was not thoroughly discussed
and analyzed in the literature. This paper studies
the trade-off between vaccine efficacy and abundance and then between
efficacy and time of availability, and the corresponding expected
outcomes for deaths and infections.

The paper is organized as follows: section 2 presents the theoretical
model, section 3 puts forward the results and discussions
about the efficacy, deployment and timing rates while section 4 concludes the paper.

\begin{figure}
\begin{centering}
\includegraphics[scale=0.4]{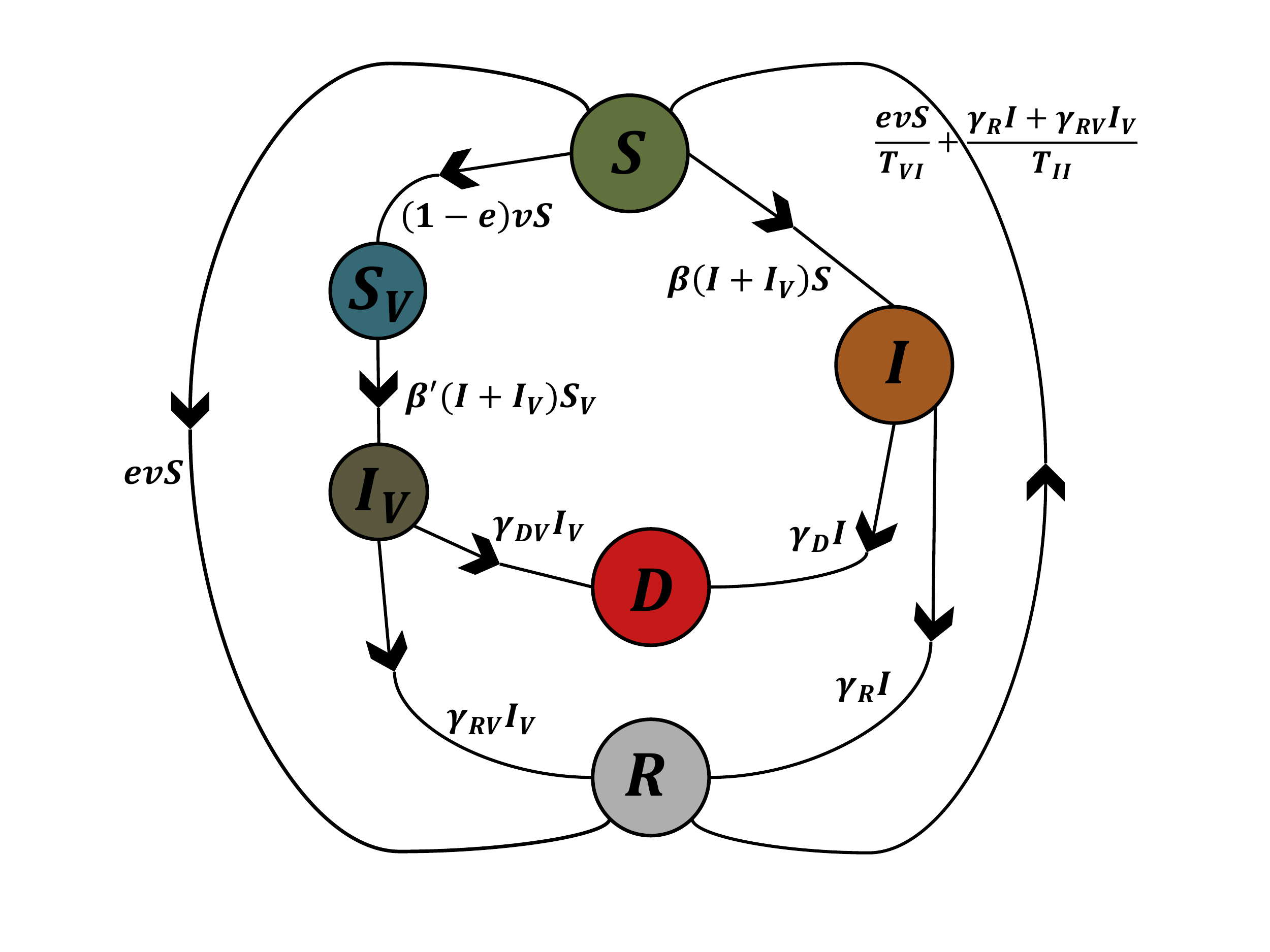}
\par\end{centering}

\caption{A schematic diagram of the $SS_{V}II_{V}RD$ model showing the six
compartments of the model (Susceptible, Vaccinated susceptible, Infectious,
Vaccinated Infectious, Recovered and Dead) together with their transfer
dynamics and the coefficients relating transfers among different compartments.
\label{fig:1}}

\end{figure}

\section{Theoretical Model}

We introduce a novel compartmental model to account for the effect
of vaccines, taking into account different vaccination deployment
rates ($v$), efficacies ($e$) under several scenarios of infection
spread rates represented by the reproductive number $(R_{t})$, starting
with different initial conditions in relation to different numbers
of infections, deaths and immune populations in different countries.
The $SS_{V}II_{V}RD$ model consists of 6 compartments: the susceptible
unvaccinated population $S$, the susceptible vaccinated population
$S_{V}$ who might become infected during the
assumed effective immunity period depending on the efficacy of the
vaccine, the infectious unvaccinated population $I$, the infectious
vaccinated population $I_{V}$, the recovered population $R$ who are either fully
protected by the vaccine during its effective protection time range
or who have already recovered from the infection and still possess
immunity, and finally the dead $D$. Each compartment is normalized
with respect to the total population so that $N=S+S_{V}+I+I_{V}+R+D=1$.

The $6$ compartments are interlinked (Figure \ref{fig:1}), and their dynamics can be modeled
by a system of coupled linear ordinary differential equations described below.

\begin{equation}
\begin{cases}
\frac{dS}{dt} & =-\beta\left(I+I_{V}\right)S-evS+\frac{evS}{T_{VI}}+\frac{\gamma_{R}I+\gamma_{RV}I_{V}}{T_{II}}\\
\frac{dS_{V}}{dt} & =\left(1-e\right)vS-\beta'\left(I+I_{V}\right)S_{V}\\
\frac{dI}{dt} & =\beta\left(I+I_{V}\right)S-\left(\gamma_{D}+\gamma_{R}\right)I\\
\frac{dI_{V}}{dt} & =\beta'\left(I+I_{V}\right)S_{V}-\left(\gamma_{DV}+\gamma_{RV}\right)I_{V}\\
\frac{dR}{dt} & =evS+\gamma_{R}I+\gamma_{RV}I_{V}-\frac{evS}{T_{VI}}-\frac{\gamma_{R}I+\gamma_{RV}I_{V}}{T_{II}}\\
\frac{dD}{dt} & =\gamma_{D}I+\gamma_{DV}I_{V}
\end{cases}\label{eq:model}
\end{equation}

The parameter $\beta$ is the incidence rate between susceptible
$(S)$ and all infectious people $(I+I_{V})$ and it can be determined
using the reproductive number of the infectious spread $R_{t}$ and
the average rate of recovery of unvaccinated people$\gamma_{R}$.
In this sense, $\beta=\gamma_{R}R_{t}$. The rate of recovery of vaccinated
people is given by $\gamma_{RV}$ and the incidence rate between
vaccinated susceptible people and infectious people is given by $\beta'=\gamma_{RV}R_{t}$.
The rate of death of unvaccinated people is $\gamma_{D}$ and that
of vaccinated ones is $\gamma_{DV}$. Finally, $T_{VI}$ represents
the average time duration of acquired vaccine immunity and $T_{II}$
represents the average duration of acquired post infection immunity.
The change in the total population is $\frac{dN}{dt}=0$, assuming
that $N$ remains constant during that period, neglecting other changes
due to natural growth, immigration, etc...

The transition dynamics between different compartments can be summarized
according to the following:
\begin{itemize}
\item The susceptible $S$ may become infected $I$ upon contact with the
infectious (vaccinated or non-vaccinated) at rate $\beta$. They
may also become recovered $R$ upon receiving a vaccine with efficacy
$e$ given at a daily rate $v$. In addition, recovered people $R$ 
would become susceptible again some time after recovery. Mainly,
recovered people due to infection would become susceptible again in
an infection immunity period $T_{II}$ and recovered people due to
vaccination would become susceptible again in a vaccine immunity period
$T_{VI}$.
\item The vaccinated susceptible population $S_{V}$ are the susceptibles
$S$ who got vaccinated but are still susceptible to infection,
and may become infectious $I_{V}$ upon a meeting rate $\beta'$
with other infectious people.
\item The infectious compartment $I$ is populated by unvaccinated susceptibles
$S$ encountering other infectious agents at a rate $\beta$, while
exit from this compartment is attributed to deaths at rate $\gamma_{D}$
into compartment $D$ and recoveries at rate $\gamma_{R}$ into compartment
$R$.
\item The vaccinated infectious compartment $I_{V}$ includes vaccinated
susceptibles $S_{V}$ catching the disease at a rate $\beta'$ and
diminished by people dying at rate $\gamma_{DV}$ into $D$ and people
recovering a rate $\gamma_{RV}$ into $R$.
\item The recovered population $R$ is formed by effectively vaccinated
susceptible people coming from $S$ and people surviving the infection
at rates $\gamma_{R}$ and $\gamma_{RV}$ from infectious and vaccinated
infectious populations $I$ and $I_{V}$. Simultaneously, recovered
people would eventually lose their acquired immunity on average periods
of $T_{II}$ after infection and $T_{VI}$ after vaccination, thus
would exit the recovered compartment $R$ back to the susceptibles
$S$.
\item Finally, the dead $D$ increase at death rates $\gamma_{D}$ and $\gamma_{DV}$
among infectious and vaccinated infectious populations $I$ and $I_{V}$. In this model, we only consider
the deaths caused by COVID-19, disregarding net population changes (natural deaths as well as natural births). 
Consequently, the dead compartment gets populated from the infected and vaccinated infected populations.
\end{itemize}
The vaccination deployment rate $v$ depends on the available supply
of the vaccine as well as the logistical capability and the popular
demand at a given time \cite{deployment}. In this paper
it is assumed to be equivalent to the daily vaccination percentage of the susceptible population,
with different scenarios representing slow, moderate and fast deployment
rates. The vaccine efficacy varies among different employed vaccines
as well as in relation to new emerging variants in addition to the
possibility of supplying a single dose of a double dosed vaccine under
short supply. The model accounts for a range of scenarios with low,
moderate and high efficacy corresponding to the former cases. We also
inspect those scenarios under different reproductive rates $R_{t}$
which depend on different values of mitigation measures related to
social distancing, protective masks, sanitation, and other factors
that alter the rate of infection spread. We assume different scenarios
of low, moderate, high and alternating (where $R_{t}$ varies in a
periodic pattern between high and low extrema) reproductive rates.

The numerical values of these parameters are taken in relation to
available data and research. The efficacy values vary between $0.5\leq e\leq0.95$
depending on the available vaccines \cite{Efficacy 1,Efficacy 2,Efficacy 3,Efficacy 4,Efficacy 5,Efficacy 6,Efficacy 7}.
The rate of recovery for vaccinated people $\gamma_{R}=\frac{1}{14}$
in relation of an average of $14$ days needed for recovery \cite{recovery},
while the death rate amounts to around $2\%$ \cite{Worldometer}
of the infected which leads to $\gamma_{D}=\frac{\gamma_{R}}{50}$.
We also assume that those who are infected after vaccination would
need a similar time of recovery, despite the fact that their symptoms
would be much reduced \cite{Sympt,VacDeath2}, thus $\gamma_{RV}=\gamma_{R}$
but their death rate, as studies reveal, would be considerably lower
by $70-85\%$ \cite{VacDeath2,VacDeath}, which is modeled through
$\gamma_{DV}=\frac{\gamma_{D}}{5}$. We take $T_{VI}=90$ days and
$T_{II}=360$ days to represent expected periods of acquired immunity
after recovery and after vaccination respectively \cite{Immunity}.
The deployment rate $v$ is assigned hypothetical values varying between
$0.1\%$ of the susceptible population per day for slow vaccination
rollout and $1.5\%$ for the highest pace of vaccination rollout.

The reproduction rate $R_{t}$ is assigned values of $0.7,1.1$ and
$1.5$ corresponding to low, medium and high reproduction rates, in
addition to a fourth scenario where $R_{t}$ alternates sinusoidally
between $0.4$ and $2$ according to the relation $R_{t}=0.8\cos(\frac{t}{15})+1.2$
where $t$ represents the time in days, with a period of around $94$
days to simulate the effect of the consecutive waves of the spread
of the infection, as observed empirically \cite{InfWave1,InfWave2}.
The periodicity of the function $\cos(at+b)$  is mathematically defined to be  $\frac{2\pi}{a}$. 
The parameter $\frac{1}{15}$ in our model simulates a periodicity of $30\pi$ 
which approximately corresponds to a $94$ day period. 
Its dimension is time given in the unit of days. 
Thus, this forecast simulates the recurrence of waves of high then low 
infection rates on an average period of 94 days. 
The growth rate of the active cases of COVID-19 infections is 
studied in the United Kingdom, South Africa and Brazil \cite{UK, Brazil}.
 These articles present the progression of the registered values of  $R_{t}$ since
 the start of the spread until recently. In both countries, $R_{t}$ assumed values
 that varied between $0.5$ up to $3$ depending on the mitigation measures and closures.
 The registered data constitutes a motivation to simulate the possible scenarios
 of infection spreads in presence of vaccines for representative values of $R_{t}$ 
corresponding to low, middle and high rates of spread.

We simulate this model under different combinations of efficacy $e$,
deployment rate $v$ and reproductive number $R_{t}$ to analyze the
corresponding cumulative numbers of infected, recovered and dead populations.

\begin{figure}
\begin{raggedright}
\includegraphics[scale=0.55]{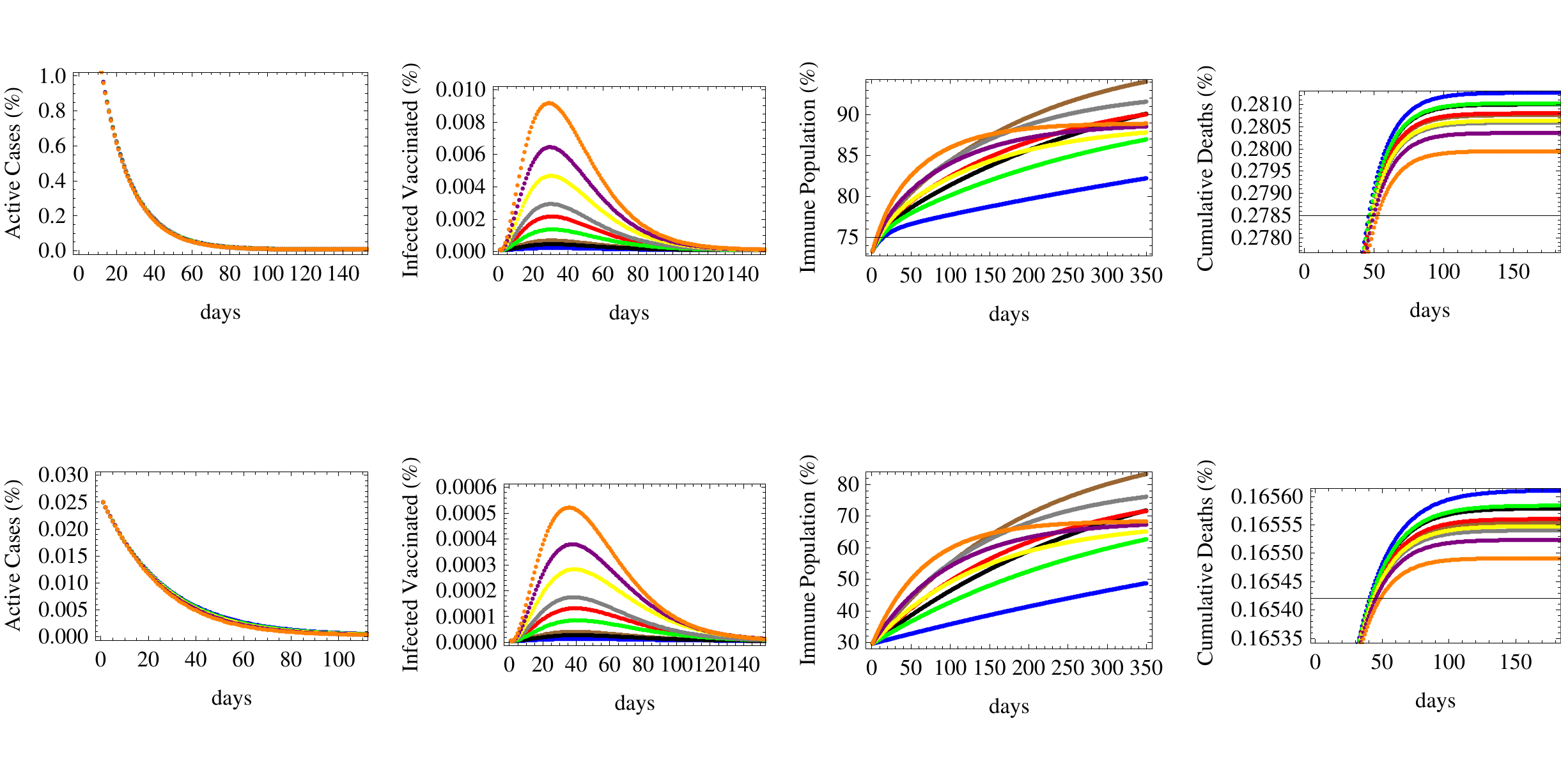}
\par\end{raggedright}

\caption{The relative numbers of active cases, infected vaccinated cases, immune
population and total cumulative deaths are shown for a low reproductive
rate $R=0.7$ under initial conditions of: $73.1\%$ vaccinated population
$2.0\%$ currently infected and $0.23\%$ dead (upper row, United Kingdom) and $29.2\%$
vaccinated population $0.02\%$ currently infected and $0.16\%$ dead
(lower row, South Africa) for nine different vaccination scenarios. The forecast
corresponds to vaccine efficacy $(e)$ and daily deployment rate $(v)$
of: $e=92\%,v=0.5\%$ (in brown), $e=92\%,v=0.3\%$ (in black), $e=92\%,v=0.1\%$
(in blue), $e=72\%,v=0.7\%$ (in gray), $e=72\%,v=0.5\%$ (in red),
$e=72\%,v=0.3\%$ (in green), $e=55\%,v=1.5\%$ (in orange), $e=55\%,v=1\%$
(in purple) and $e=55\%,v=0.7\%$ (in yellow). \label{fig:2}}
\end{figure}

\section{Results and discussions}

\subsection{Efficacy vs abundance}

\begin{figure}
\begin{raggedright}
\includegraphics[scale=0.55]{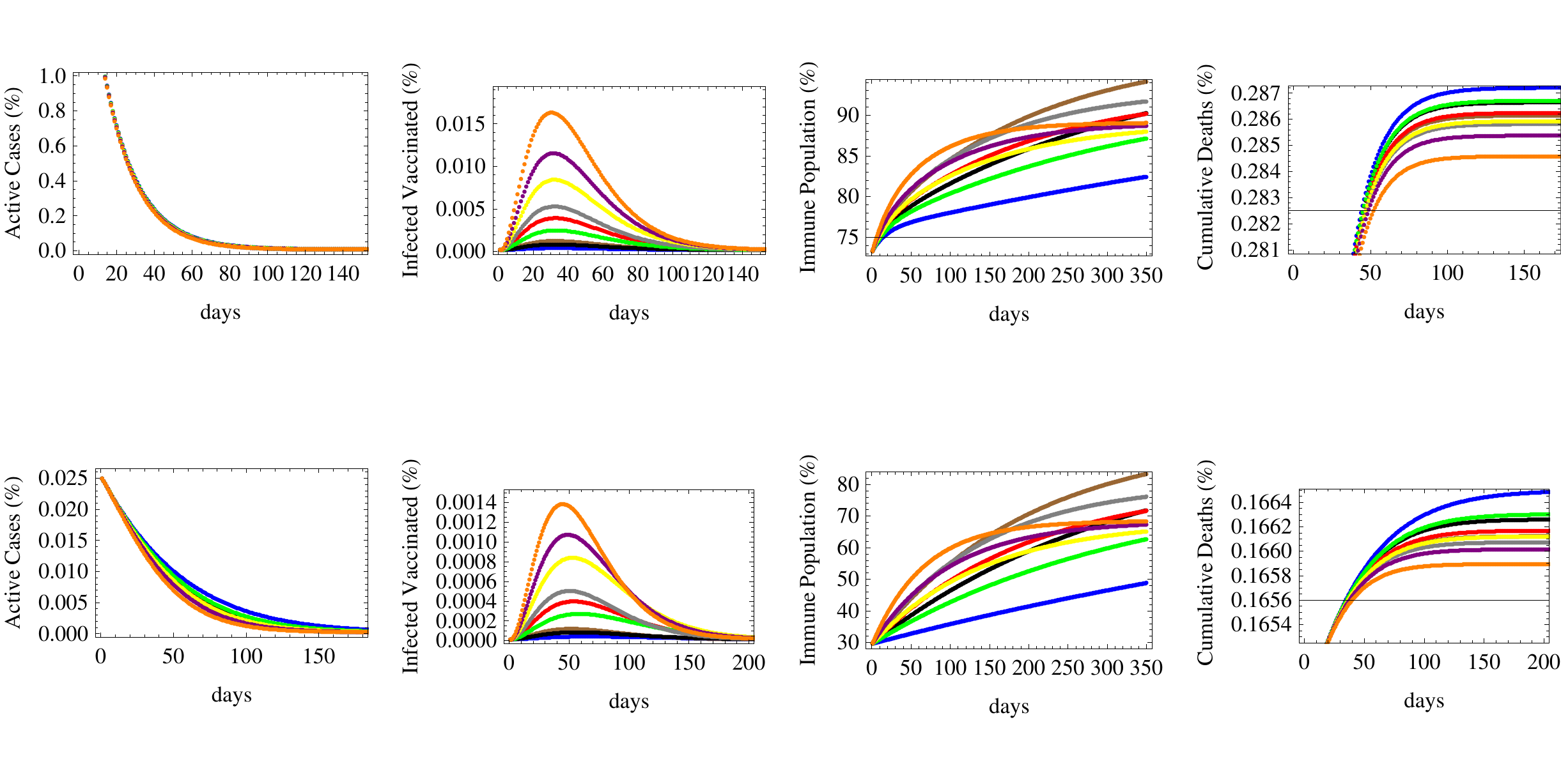}
\par\end{raggedright}

\caption{The figure shows the relative numbers of active cases, infected vaccinated
cases, immune population and total cumulative deaths for a medium
reproductive rate $R=1.1$ with the same two initial conditions and
nine vaccination scenarios of Figure \ref{fig:2}. \label{fig:3}}
\end{figure}

\begin{figure}
\begin{raggedright}
\includegraphics[scale=0.55]{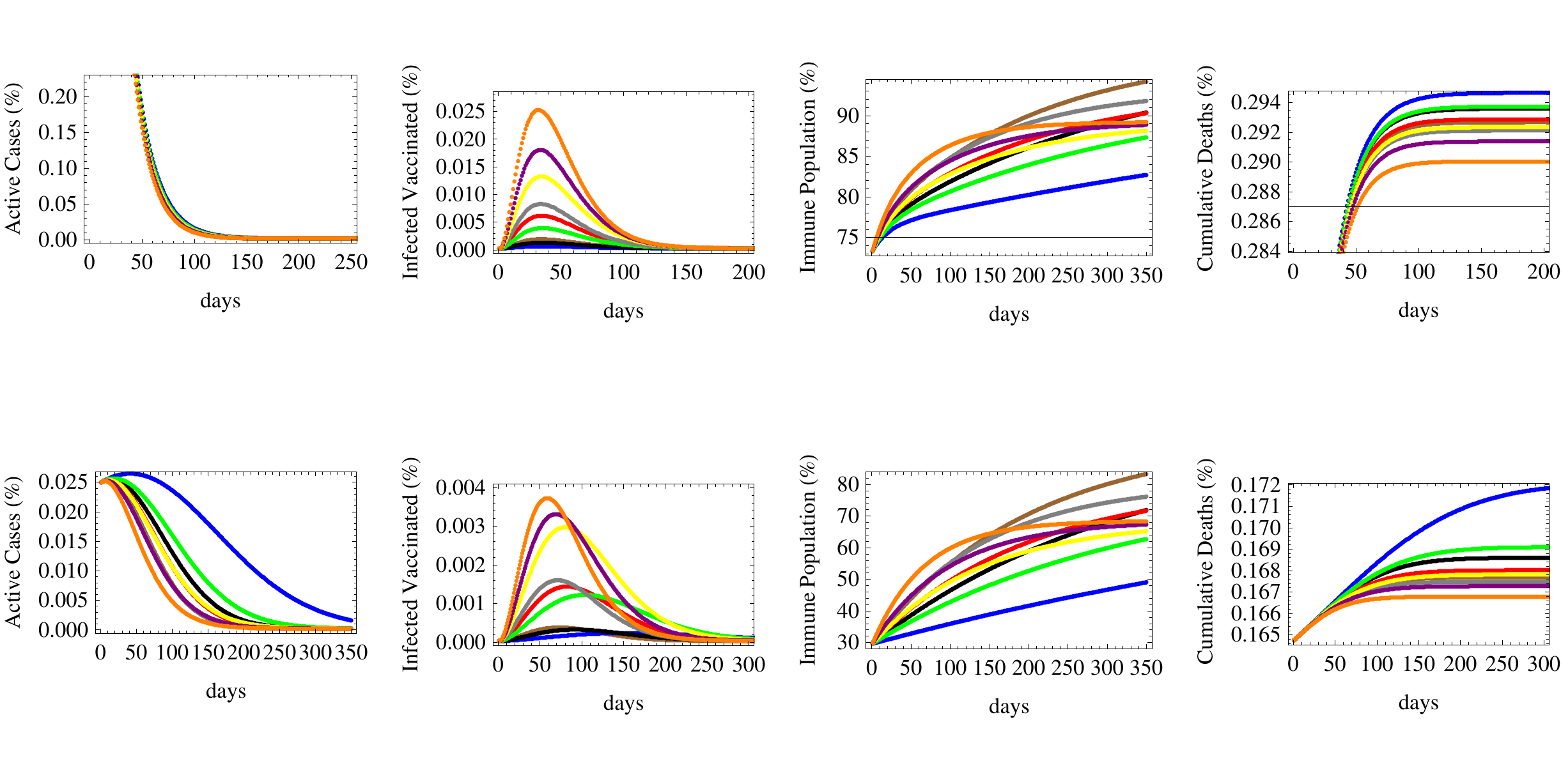}
\par\end{raggedright}

\caption{The figure shows the relative numbers of active cases, infected vaccinated
cases, immune population and total cumulative deaths for a high reproductive
rate $R=1.5$ with the same two initial conditions and nine vaccination
scenarios of Figure \ref{fig:2}. \label{fig:4}}
\end{figure}

\begin{figure}
\begin{raggedright}
\includegraphics[scale=0.55]{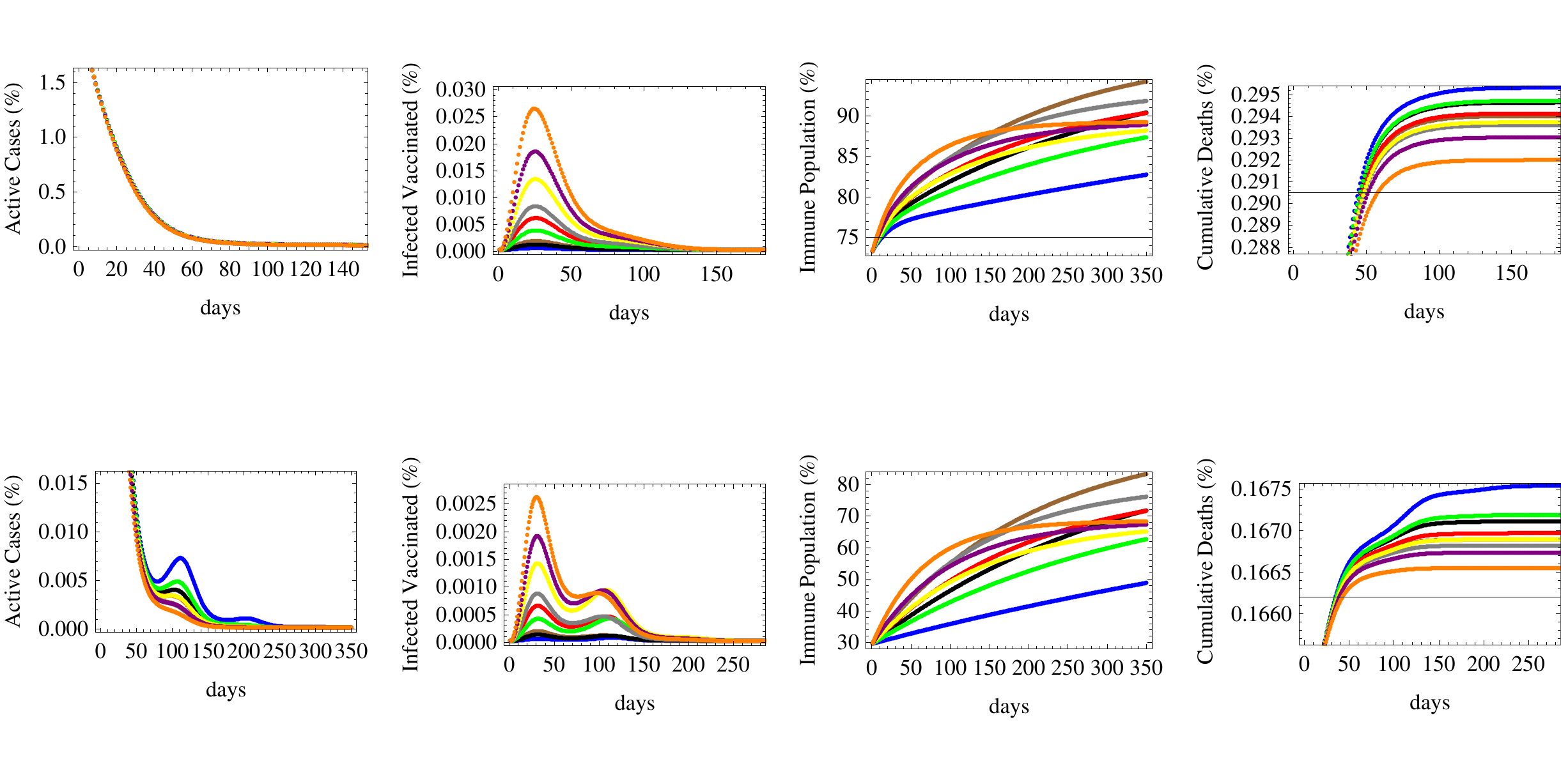}
\par\end{raggedright}

\caption{The figure shows the relative numbers of active cases, infected vaccinated
cases, immune population and total cumulative deaths for an alternating
reproductive rate, varying sinusoidally between $R=0.4$ and $R=2$,
with the same two initial conditions and nine vaccination scenarios
of Figure \ref{fig:2}. \label{fig:5}}
\end{figure}

We simulate the theoretical model introduced in (\ref{eq:model})
to determine relative numbers of active cases, infected vaccinated
cases, immune population and total cumulative deaths under nine different
combinations of efficacies $(e)$ of available vaccines and their
deployment rates $(v)$ given by: $e=92\%,v=0.5\%$, $e=92\%,v=0.3\%$,
$e=92\%,v=0.1\%$, $e=72\%,v=0.7\%$, $e=72\%,v=0.5\%$, $e=72\%,v=0.3\%$,
$e=55\%,v=1.5\%$, $e=55\%,v=1\%$ and $e=55\%,v=0.7\%$ in order
to compare the levels of infection, death and immunity between higher
efficacy vaccines at lower abundance and lower efficacy vaccines with
more abundance or deployment rate, together with middle values between
them. 
\paragraph{Case Study}
We simulate in this paper the actual data of infections, deaths and vaccination
in two countries with high vaccination rate (United Kingdom) and low vaccination
rate (South Africa), according to data publically available in \cite{Worldometer}.
The initial conditions corresponding to the United Kingdom are: $73.1\%$ vaccinated population
$2.0\%$ currently infected and $0.23\%$ dead, while those of South Africa are:$29.2\%$
vaccinated population $0.02\%$ currently infected and $0.16\%$ dead. 
In this sense, the results of this forecast are not country specific but are of global
significance. They can be applied to any other country with similar conditions.
 We repeat this simulation for different levels of infection
spread modeled through low, medium, high and alternating reproductive
rates $R_{t}$ defined before. The results are displayed in Figures
\ref{fig:2},\ref{fig:3},\ref{fig:4} and \ref{fig:5}, respectively.

Figure \ref{fig:2} shows that when the reproductive rate is low
($R_{t}=0.7$), the number of active cases would fall down quickly
under all vaccination schemes, for various efficacies and deployment
rate, both in countries that are in the middle or early stages of
the spread (upper and lower figures respectively). The number of infected
vaccinated people is the highest for the low efficacy vaccine. Herd
immunity would be achieved optimally under fast deployment of the
high efficacy vaccine and the middle efficacy vaccine, while the lowest
immunity is achieved under the slowest deployment of the high efficacy
vaccine. This shows that the deployment rate is an essential factor
in combination with efficacy for achieving immunity. The figure also
shows that the number of deaths is most reduced under the two fastest
deployment schemes of the low efficacy vaccine, followed by the fastest
deployment rate of that of middle efficacy. The latter result shows
that to reduce deaths, the most essential measure is to ensure the
the availability and the fast deployment of the vaccine, while reaching
herd immunity and decreasing infections (thus lowering the pressure
on the health sector and decreasing the possibility of appearance
of new mutations) depends relatively more on its efficacy.

In the case of a medium reproductive rate ($R_{t}=1.1$) corresponding
to a slow increase in the number of infections in absence of vaccine,
Figure \ref{fig:3} shows that the number of active cases would
decrease in countries that are in their middle stages of infection,
while they would keep on increasing for a limited interval of time
in countries that are in their early stages of spread. The duration
needed for the active cases to start decreasing is shortest for the
vaccine with fastest deployment rate (despite its low efficacy), while
that with the slowest deployment rate (with high efficacy) needs more
time to decrease the number of active cases. The highest number of
infected vaccinated people would be obtained through the rapidly deployed
low efficacy for countries in their middle stages of spread, while
for early stage countries, each scenario of deployment of the low
efficacy vaccine would lead to maximal infections under different
times. In a similar manner to what happens for low reproductive rate,
immunity will be maximally obtained under the deployment of high efficacy
vaccines at high rates, while the least deaths are achieved under
the fastest deployment rates of vaccines.

For a country under a fast spread scenario simulated by $R_{t}=1.5$,
Figure \ref{fig:4} shows that the number of active cases would
fall in a middle spread stage country, though it takes more time than
that with lower $R_{t}$. However, in early spread countries, the
number of active cases will rise and peak after a period of time before
falling down after a relatively long time (around $100-200$ days)
of vaccination. The slowest vaccination schemes of the low and middle
efficacy vaccines would reach the highest peaks and need more time
for controlling the spread while the fastest vaccination schemes would
achieve it the quickest. The number of infected vaccinated people
would be the largest in the case of the low efficacy vaccine deployed
rapidly in a middle spread stage country, while for a country at early
spread stages, the two slowest deployments of the low efficacy vaccine
lead to the highest infections among the vaccinated population. Immunity
will be maximally obtained under the deployment of high efficacy vaccines
at high rates, while the least deaths are achieved under the fastest
deployment rates of vaccines, like what was shown before for low and
medium spread rates, but with a much higher magnitude of deaths in
all scenarios due to a higher reproductive rate.

In the more realistic case of an alternating reproductive number varying
between $0.4\leq R_{t}\leq2$ and corresponding to consecutive patterns
of low and high spread waves, we realize that in a country at a middle
spread stage, all vaccination strategies would lead to bringing down
the number of active cases with a small peak arising after the return
of the next wave, while in a country at early stages of spread, the
slow deployment of vaccines, whether with high or low efficacies,
would cause the re-emergence of several peaks of infections and high
number of active cases. Only the fast deployment of vaccines of various
efficacies would lead to curb the number of infections just after
the first peak of spread. The number of infected vaccinated people
would be the highest for the case of fast deployment of low efficacy
vaccines in countries with middle stages of spread while stronger
peaks of infections among the vaccinated people would occur under
slower vaccination scenarios of low efficacy vaccine during the second
wave in countries with early spread stages. The highest herd immunity
levels would be attained under the scenarios of fastest deployments
of high and medium efficacy vaccines in both categories of countries
under early or middle stages of spread after $3-4$ months, despite
the early lead and the sharp rise in the immune population during
the first few weeks of fast deployment of low efficacy vaccines. The
cumulative number of deaths would rise in jumps corresponding to successive
waves of spread especially in early stage countries, but in both cases,
the fastest vaccination schemes of low efficacy vaccines would ultimately
lead to the minimal number of deaths while the slowest pattern of
high efficacy vaccine deployment would result in the highest cumulative
rate of deaths. Those simulations are displayed in Figure \ref{fig:5}.

\begin{figure}
\begin{raggedright}
\includegraphics[scale=0.5]{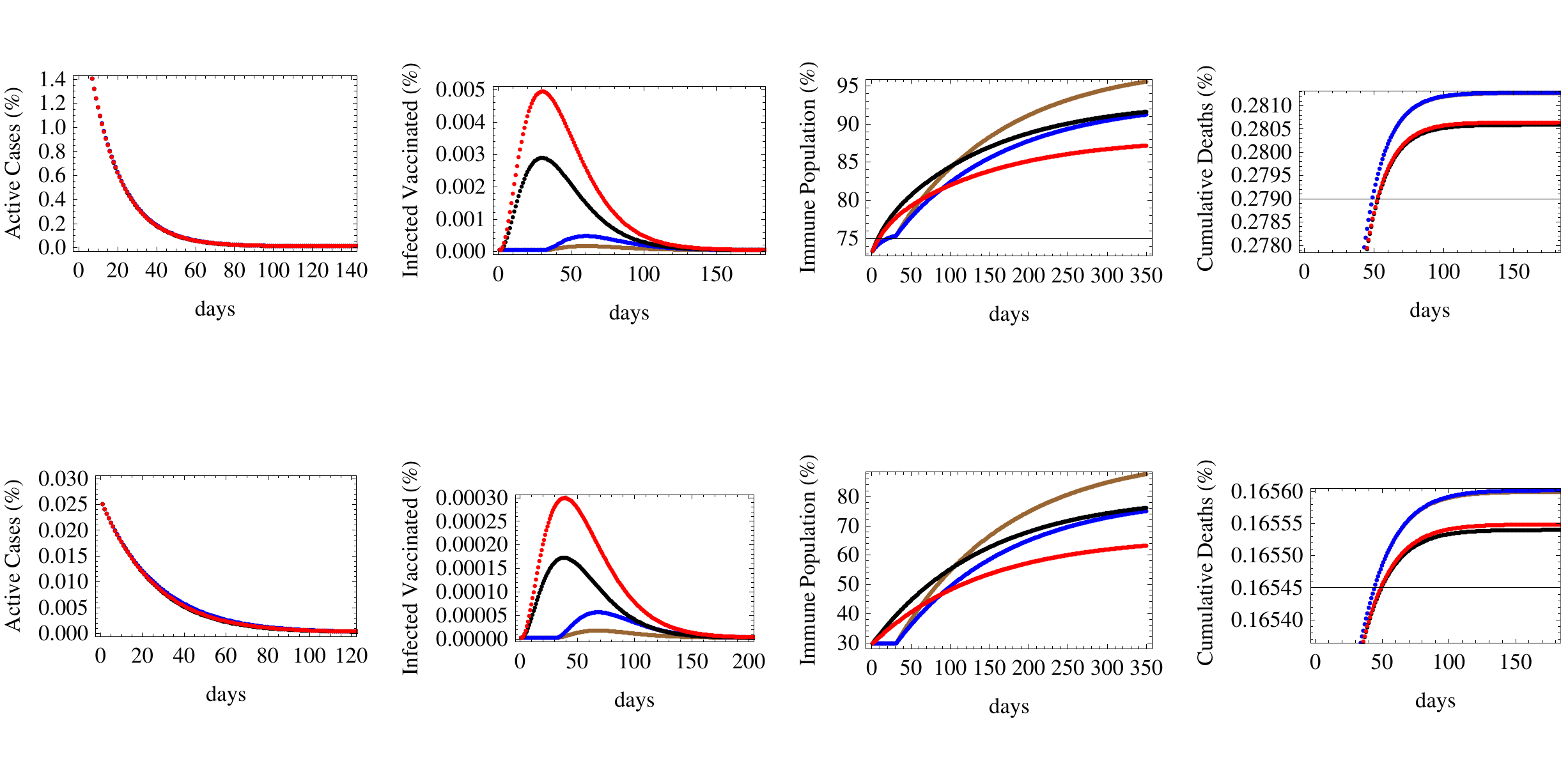}
\par\end{raggedright}

\centering{}\caption{The figure shows the expected relative numbers of active cases, infected
vaccinated cases, immune population and cumulative deaths at a low
reproductive rate $R=0.7$ in relation to four cases simulating the
effect of efficacy and the time of initiation of the vaccination process:
a vaccine of efficacy $e=52\%$ to start deployment immediately (red),
a vaccine of efficacy $e=72\%$ to start deployment immediately (black),
a vaccine of efficacy $e=72\%$ to start deployment after 30 days
(blue) and a vaccine of efficacy $e=92\%$ to start deployment after
30 days (brown), all being deployed at an equal rate of $0.7\%$ of
the susceptible population per day. The upper row corresponds to initial
conditions of $40.6\%$ vaccinated population $1.8\%$ currently infected
and $0.18\%$ dead while the lower row corresponds to $3.1\%$ vaccinated
population $0.46\%$ currently infected and $0.11\%$ dead. \label{fig:6}}
\end{figure}

\subsection{Efficacy vs time of availability}

\begin{figure}
\begin{raggedright}
\includegraphics[scale=0.55]{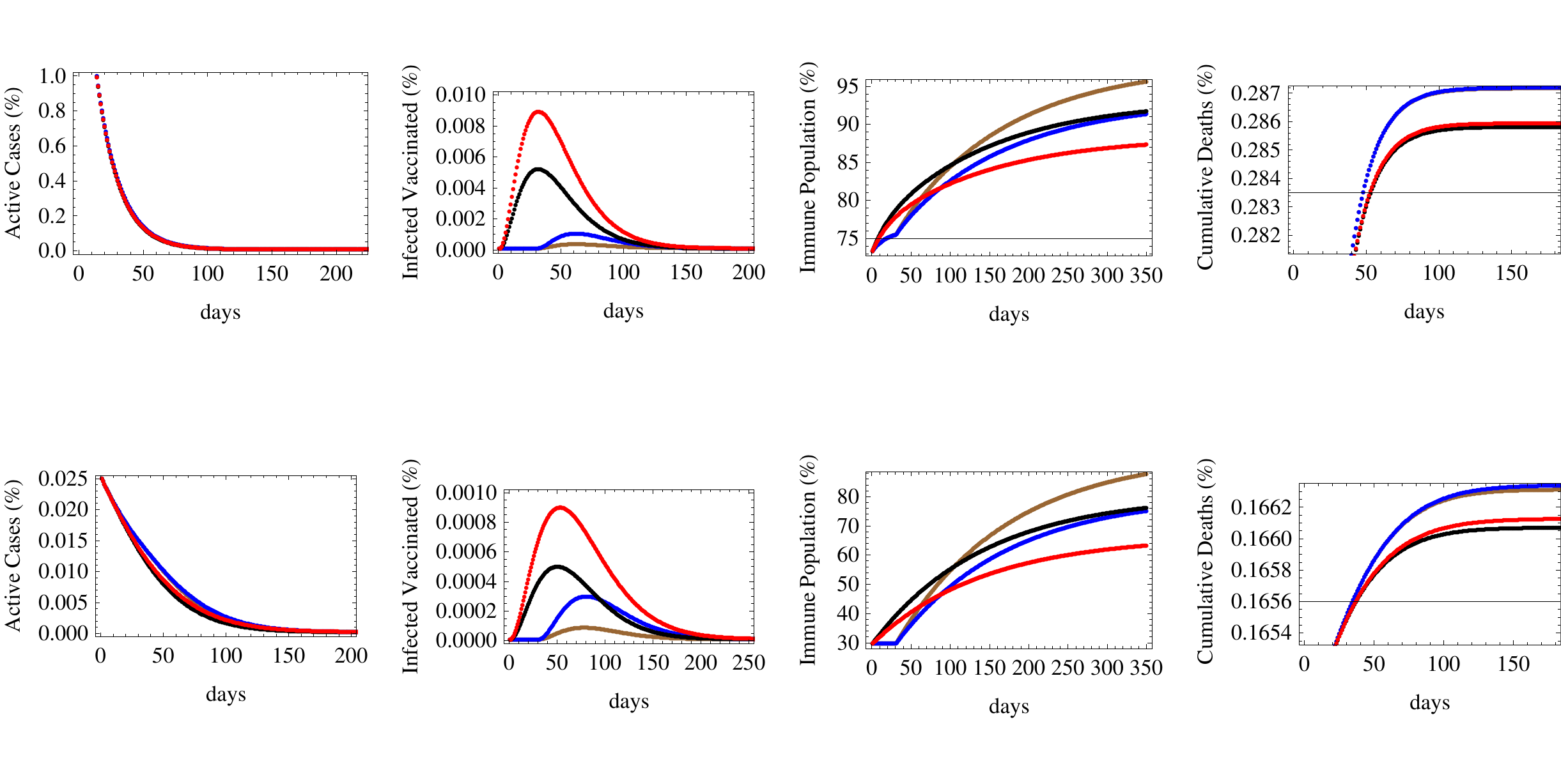}
\par\end{raggedright}

\caption{The figure shows the expected relative numbers of active cases, infected
vaccinated cases, immune population and cumulative deaths corresponding
to a medium reproductive rate $R=1.1$ with same initial conditions,
efficacies and time of initiation as those detailed in Figure \ref{fig:6}.
\label{fig:7}}
\end{figure}

\begin{figure}
\begin{raggedright}
\includegraphics[scale=0.55]{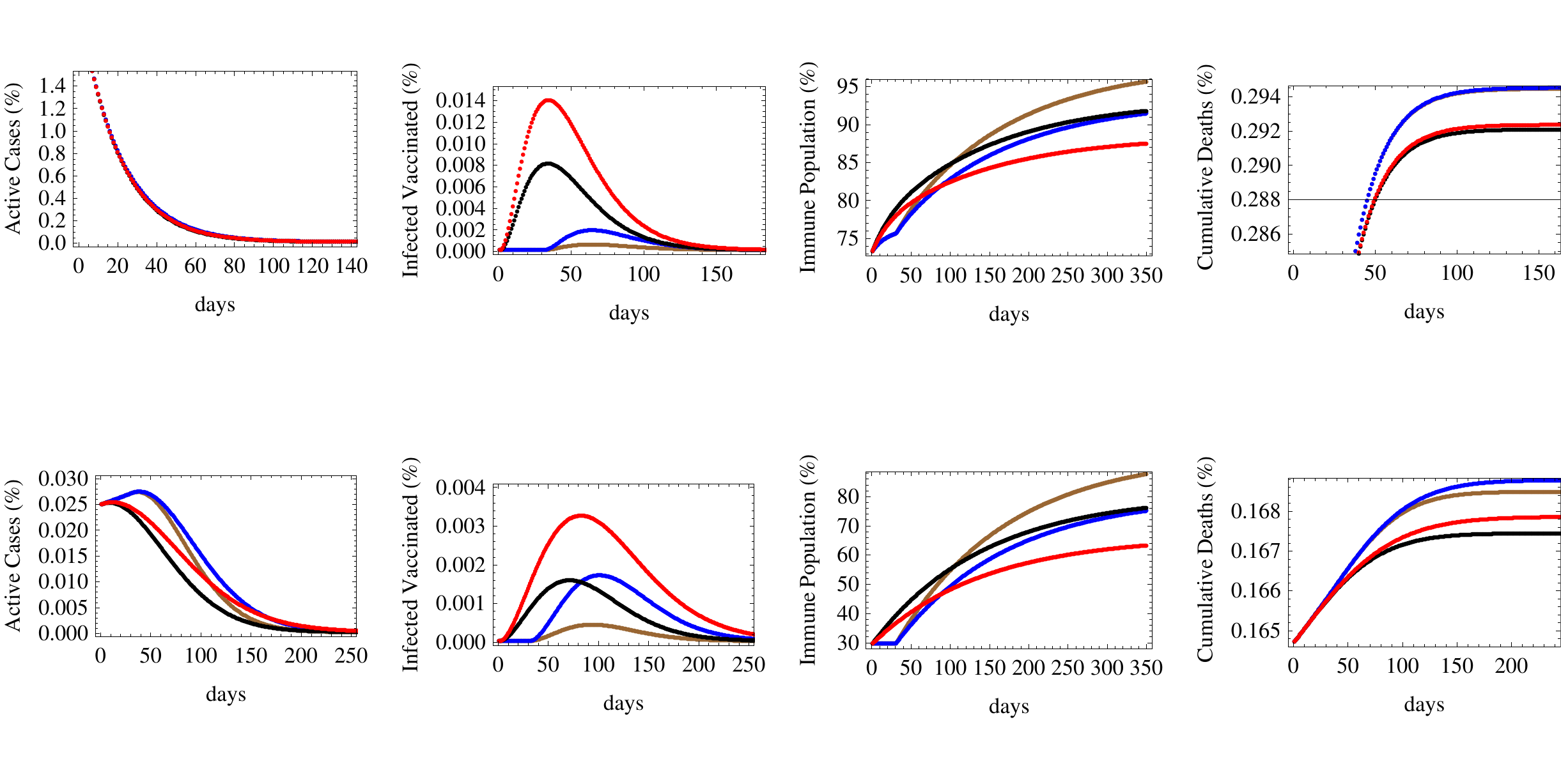}
\par\end{raggedright}

\caption{The figure shows the expected relative numbers of active cases, infected
vaccinated cases, immune population and cumulative deaths corresponding
to a high reproductive rate $R=1.5$ with same initial conditions,
efficacies and time of initiation as those detailed in Figure \ref{fig:6}.
\label{fig:8}}
\end{figure}

An important aspect of vaccination strategies that we consider in
this paper is the question of vaccine efficacy versus time of availability
or start of deployment. In this simulation, we forecast the relative
number of active cases, infected vaccinated people, the immune population
and the cumulative number of deaths under low, medium and high reproductive
rates $R=0.7,1.1$ and $1.5$ respectively in Figures \ref{fig:6},
\ref{fig:7} and \ref{fig:8}. Four cases simulating the effect of
efficacy and the time of initiation of the vaccination process were
considered: a vaccine of low efficacy $e=52\%$ to start deployment
immediately (red), a vaccine of medium efficacy $e=72\%$ to start
deployment immediately (black), a medium efficacy vaccine with $e=72\%$
to start deployment after 30 days (blue) and a vaccine of high efficacy
$e=92\%$ to start deployment after 30 days (brown), all being deployed
at an equal rate of the susceptible population per day. As in previous
simulations, the upper row corresponds to countries at middle stages
of spread with initial conditions of $40.6\%$ vaccinated population
$1.8\%$ currently infected and $0.18\%$ dead while the lower row
corresponds to countries at early stages with $3.1\%$ vaccinated
population $0.46\%$ currently infected and $0.11\%$ dead.

Under the circumstances of low reproductive rate $R_{t}=0.7$ depicted
in Figure \ref{fig:6}, it is clear that the number of active cases
will fall significantly under the four efficacy-timing schemes in
both early and middle spread stage categories. The number of infected
vaccinated people in both categories would occur under the scenario
of immediate deployment of low efficacy vaccines. Herd immunity would
be maximally attained through the adoption of high efficacy vaccines
being deployed with a one month period of delay while it would be
the lowest using low efficacy vaccines deployed immediately. On the
level of cumulative deaths, the least number of deaths is realized
under the scenario of immediate deployment of medium then low efficacy
vaccines, while the delay would raise the number of deaths even while
using high efficacy vaccines.

Figure \ref{fig:7} simulates same efficacy-timing scenarios under
medium reproductive rate $R_{t}=1.1$. We notice that in countries
in their middle stage of infection spread, the number of active cases
would fall quickly under all scenarios, while in early stage countries,
the late deployment scenarios would cause a rise and a peak in infections
before falling down significantly, whereas the quick deployment of
low or medium efficacy vaccines would lower the active cases faster
during the first few months. We also notice that in about $100$ days,
the number of active cases due to a delayed high efficacy vaccine
will catch up and fall below the expected active cases under an immediately
deployed low efficacy vaccine. The number of infected vaccinated people
is the highest for the lowest efficacy vaccine and the lowest for
the highest efficacy vaccine in both country categories. Similarly,
for both categories, herd immunity is maximally attained using the
delayed high efficacy vaccine rather than the immediate low efficacy
one which provides the lowest percentile of immune population. However,
regarding deaths, the lowest numbers of deaths are attributed to the
immediate deployment of medium then low efficacy vaccines, while a
delayed deployment will cause more deaths even while using high and
medium efficacy vaccines.

Under the scenario of high reproductive number simulated by $R_{t}=1.5$,
the number of active cases would fall slowly in countries that are
in their middle stages of vaccination and spread, while it would rise
and peak under all vaccination scenarios in early stage countries.
The highest peaks are attributed to delayed vaccinations even though
infections would fall rapidly once vaccination starts, and eventually
the corresponding number of active cases would fall below that of
low efficacy vaccine being deployed immediately. Immediate deployment
also helps to flatten the curve on infections, thus reducing the expected
peak number of cases. The maximum number of infected vaccinated people
would correspond low efficacy vaccine deployed immediately while the
lowest corresponds to the high efficacy vaccine deployed late, for
both country categories. Similarly, herd immunity is maximally achieved
by the high efficacy vaccine despite being introduced late while the
lowest level of immunity is caused by the low efficacy vaccine despite
early introduction. The immediate deployment of medium efficacy vaccine
minimizes the number of deaths, followed by the immediate low efficacy
vaccine, while late deployment raises the number of deaths for both
country categories, but with higher relative differences among the
death outcomes for middle stage countries.

\subsection{Limitations}

This study takes into account all parameters related to vaccine efficacy
and deployment rates under several infectious rates and initial conditions.
However, there is a limiting factor related to the maximal proportion
of the population who are willing or at least who would eventually
take the vaccine. Here, we took this into account indirectly by using
a daily deployment rate $v$ which is proportional to the susceptibles
$S$, and not to the total population $N$. In this sense, in the
initial phases of vaccination, the number of people taking the vaccine
would be the highest, but as time progresses, the number of susceptibles
decreases, hence the daily number of vaccinated people decreases.
It is natural to assume that, as when a country reaches a high level
of vaccination, less people will be willing to get the vaccine. If
vaccination rates were only connected to abundance or logistic infrastructure,
they would have been linked to the total population $N$.
It is also important to mention that the reproduction number $R_{t}$ 
can differ for different population groups. Its value is dominated by the group in which most transmission occurs.
In this sense, it is not a homogeneous parameter that applies on a nationwide or social scale but depends
on demographic, socio-economic and other factors.

Due to various reasons ranging from religious and political beliefs,
into non-scientific and anti-vaxxer fears, there might be a sizable
sector of the society who would refuse to get vaccinated \cite{Asymmetric}.
Vaccine hesitancy is not directly simulated in the model, but it is
indirectly represented through relating the daily deployment rate
to the number of susceptibles hence it decreases as the number of
vaccinated people increases. In our scenario, immunity of this portion
of the population would still be achieved through infection rather
than vaccination.

\section{Conclusion}

In this paper we introduced a general novel compartmental model accounting
for the vaccinated population, infected vaccinated population, active
infections, and deaths with various vaccine efficacies and vaccination
deployment rates.

We simulated different scenarios and initial conditions, and we showed
that abundance and higher rate of deployment of low efficacy vaccines
would lower the cumulative number of deaths in comparison to slower
deployment of high efficacy vaccines. However, the high efficacy vaccines
can better lower the number of active cases and achieve faster and
higher herd immunity.

We also discovered that at the same daily deployment rate, the earlier
introduction of vaccines with lower efficacy would also lower the
number of deaths with respect to a delayed introduction of high efficacy
vaccines, which can, however, lower the number of infections and attain
higher levels of herd immunity.

\end{document}